\documentclass[aps,prl,twocolumn,showpacs,floatfix]{revtex4}
\usepackage{graphicx}

\begin{document}

\title{Role of disorder in the size-scaling of material strength}

\author{Mikko J. Alava$^{1}$}
\affiliation{$^{1}$ Laboratory of Physics, Helsinki University of Technology, FIN-02015 HUT}
\author{Phani K. V. V. Nukala$^2$}
     \affiliation{$^2$ Computer Science and Mathematic Division,
Oak Ridge National Laboratory, Oak Ridge, TN 37831-6359, USA}
\author{Stefano Zapperi$^{3,4}$}
    \affiliation{$^3$ INFM-CNR, SMC, Dipartimento di Fisica,
    Universit\`a "La Sapienza", P.le A. Moro 2
        00185 Roma, Italy }
\affiliation{$^4$ ISI Foundation, Viale S. Severo 65, 10133 Torino, Italy }

\begin{abstract}
We study the sample size dependence of the strength of
disordered materials with a flaw, by 
numerical simulations of lattice models for fracture.
We find a crossover between a regime controlled by the
fluctuations due to disorder and another controlled by
stress-concentrations, ruled by continuum fracture mechanics. 
The results are formulated in terms of a scaling law involving 
a statistical fracture process zone. Its existence and scaling
properties are only revealed by 
sampling over many  configurations of the disorder. 
The scaling law is in good agreement with experimental results obtained 
from notched paper samples.
\end{abstract}


\maketitle

\date{\today}

The fracture strength of materials depends 
on various characteristic length-scales of the specimen, and
represents a fundamental open problem of science and engineering.
Probably the oldest scientific study of this issue was performed by Leonardo da
Vinci, who measured the carrying-capacity of metal wires of varying
length \cite{leonardo40}. The simple observation was that the longer
the wire, the less weight it could sustain. The reason for
this behavior is rooted in the structural disorder present
in the material: the strength is dominated by the weakest part
of the sample and its distribution is related to
extreme value statistics \cite{gumbel}. Longer wires
are more likely to contain weak parts and are thus bound
to fail earlier on average. While the physical mechanism
behind this extreme-value based statistical size effect is clear,
obtaining mathematical laws for it is still a formidable task.
In quasi-brittle materials, such as concrete and many other composites,
this issue is particularly important and it is complicated 
by the significant damage accumulation preceding sample failure.

The most important setting to study size effects 
involves a specimen containing a pre-existing flaw, {\it a notch}. 
Failure in this case is determined
by the competition between deterministic effects, due to the stress
enhancement created by notch, and the stochasticity coming from the
disorder \cite{bazantbook}. For a sufficiently large notch,
stress enhancement around the crack tip dominates the process.
It is then customary to consider disorder as a small perturbation,
by defining a Fracture Process Zone (FPZ) around the crack tip, where all the
damage accumulation is confined. For quasi-brittle materials, however, the
size of the FPZ may not be negligible when compared to the system size.
Conversely, for small notches failure is influenced by
statistical effects, and may for instance initiate far from the
pre-existing notch due to nucleated microcracks. 
Several formulations to account for the size
effects have been proposed in the literature
\cite{bazant04b,hu92,karihaloo99,morel00,morel02,carpinteri05}.
These approaches are mainly based on the ad hoc extensions
of linear elastic fracture
mechanics (LEFM). This is a well established framework to understand cracks
in homogeneous media but encounters fundamental problems when 
disorder is strong and homogenization methods are not applicable.

In LEFM the stability of a flaw against failure is given by the
Griffith's energy criterion for the critical stress, $\sigma_c \sim
K_c/\sqrt{a_0}$, where $a_0$ is the linear size of the crack
and the critical stress intensity factor $K_c \sim \sqrt{E G_c}$ is a
function of the fracture toughness $G_c$ and the elastic modulus $E$
\cite{griffith20}. A scaling law for the
size-effect has been proposed by Bazant for
quasi-brittle materials \cite{bazant04b}. One
generalizes the Griffith expression by postulating an additional
length-scale $\xi$  due to the presence of a FPZ
\begin{equation}
\sigma_c = K_c/\sqrt{\xi + a_0}.
\label{baz}
\end{equation}
Equation~(\ref{baz}) tries to incorporate two natural,
important effects: First, in the large notch
limit $\xi/a_0 \ll 0$ one should recover an expression that follows
the LEFM scaling, in which the strength is inversely proportional to
$1/\sqrt{a_0}$. Second, for a vanishing external flaw size
$a_0\rightarrow0$, the average strength should still remain finite.

Here we investigate the role of the disorder in the failure of notched
quasi-brittle specimens, providing a microscopic justification and
establishing the limits of validity of Eq. (\ref{baz}).
We study the size scaling of strength by the extensive numerical
simulations, a difficult task due to the different length scales
involved and to the need of significant statistical averaging. We
vary the disorder, which we model as a locally varying random
failure threshold, and show that it plays a crucial role in determining the size effect, 
influencing the fracture toughness $K_c$. Furthermore, the lengthscale $\xi$ 
naturally emerges from the simulations and can be shown to be 
directly related to the FPZ size. Finally, for notch sizes smaller than
a critical length $a_c$, we observe a cross-over
to the inherent, sample-size dependent strength of the unnotched sample.
We present a scaling formula that incorporates all these effects
and confirm its validity by comparing the simulations to experiments
on notched paper samples.

To simulate a disordered elastic solid, we consider the simplest case where disorder 
and LEFM-like stress enhancements can be incorporated. We perform
extensive simulations of the random fuse
model (RFM) \cite{deArcangelis85,alava06}. The RFM represents a 
quasi-brittle failure process by
an electrical analogue composed of a network of fuses. We consider a
triangular lattice of linear size $L$ with a central notch of length
$a_0$. The fuses have unit conductance (which would correspond to $E=1$
in the elastic system) and 
random breaking thresholds
$i_c$.  These represent a locally varying fracture
toughness/strength. The $i_c$ lie between 0 and 1, with a cumulative
distribution $P(i_c) = i_c^{1/D}$, where $D$ represents a
quantitative measure of disorder. The larger $D$ is, the stronger the
disorder. In the simulation, the burning of a fuse occurs irreversibly, 
whenever the electrical current
in the fuse exceeds its threshold $i_c$. An external current is
increased applying a voltage difference between the top and the
bottom lattice bus bars and applying periodic boundary conditions
along the other direction. Damage accumulates until a connected
fracture path disconnects the network and one can define the strength
$\sigma_c$ as the total peak current divided by the length of the bus bar.
We also perform numerical simulations for the random spring model (RSM)
\cite{nukala05}, similar to the mesoscale models used routinely
for concrete \cite{lilliu03}. The RSM is similar to the RFM but the fuses are
replaced by elastic springs that break when their elongation reaches
a random threshold. While the RSM represents more faithfully the 
elastic continuum, the statistical properties of the fracture
process are analogous to those of the RFM \cite{nukala05}.

\begin{figure}[ht]
\begin{center}
\includegraphics[width=8cm]{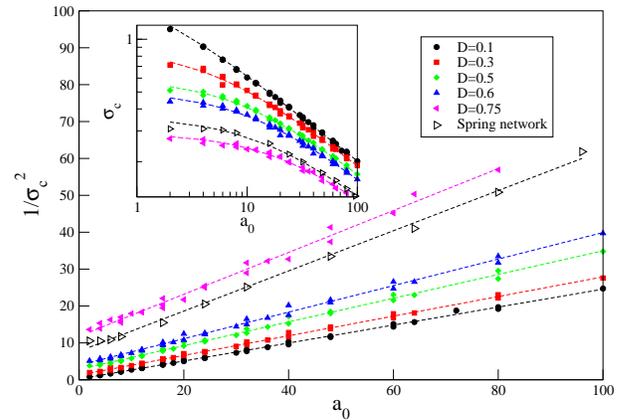}
\end{center}
\caption{Inset: The strength $\sigma_c$ as a function of the notch size $a_0$
for several disorders $D$ and system sizes $L$. 
The main figure shows the scaling plot of the data according to Eq.~(\protect\ref{baz}).}
\label{fig2}
\end{figure}

The inset of Fig.~\ref{fig2} reports the strength, averaged over different
configurations, with varying $a_0$, $D$, and $L$.  The most instructive way of plotting is to consider
the inverted square strength, $1/\sigma_c^2$. Assuming 
Eq.~(\ref{baz}), it is clear that $1/\sigma_c^2$ should become a linear
function of $a_0$ for large enough notches. Plotting the data in
such a manner in Fig.~\ref{fig2} reveals four interesting features:
({\em i}) for $a_0 \gg 1$, the scaling of Eq.~(\ref{baz}) is
recovered asymptotically; ({\em ii}) the linear part of the data when
extrapolated towards $a_0 = 0$ reveals a
disorder-dependent intercept $\xi(D)$, that should be related to the
size of the FPZ;
({\em iii}) the slope of the linear part of the data
$(1/K_c^2(D))$ is disorder-dependent, which implies a disorder-dependent
fracture toughness $G_c(D)$;
and finally, ({\em iv}) a careful observation (see Fig.~\ref{fig3}) reveals that for small
$a_0$ less than a critical crack size $a_c$, the strength scaling crosses-over from a 
stress concentration dominated LEFM scaling (Eq.~\ref{baz}) to a disorder dominated scaling. 
In particular, the data presented in the inset of Fig.~\ref{fig3} indicates that the strength of
the unnotched system (for $a_0 =0$) is smaller than the LEFM limit
$K_c/\sqrt{\xi}$ (where $1/K_c^2$ is the slope and $\xi$ is the 
intercept of the lines in Fig.~\ref{fig2} based on Eq.~(\ref{baz})). 
Hence, in order to extrapolate the typical
sample-size dependent strength from notched experiments it is necessary to
know $a_c$. As can be seen, these features are also exhibited by RSM results.

The cross-over at a critical crack size $a_c$ marks the important
role of structural disorder or internal damage on the size-effect,
and relates to the size-effect in unnotched samples. 
For the RFM, the size-effect without a notch
has been shown by simulations and theoretical arguments
to have a logarithmic dependence on the linear system size, $L$
\cite{duxbury86,alava06}. For many engineering materials, one
resorts to the Weibull theory \cite{weibull39} as an empirical
starting point. 

This cross-over from LEFM dominated strength scaling to disorder dominated 
strength scaling is illustrated in
Fig.~\ref{fig4}, where we compare the simulation data to
experimental results on paper samples with varying center notch
sizes. The paper data is from Ref.~\cite{wathen03}. The two data sets
presented are for strips of $L=15$cm cut from laboratory-made handsheets, of fine paper-type,
with center defects of nominal sizes of $a_0=$0.5, 1, 1.5, 2, and 2.5 mm.
Tensile tests were performed on 100 samples, in order to average the results. 
In Fig.~\ref{fig4}, we observe clearly the presence of $a_c$, 
below which the notches have a negligible role on $\sigma_c$, 
and for crack sizes $a_0$ larger than $a_c$ fracture is ruled by LEFM. We
have also fitted the simulation data to these experiments, by
considering similar $a_0/L$ ratios and varying the $D$ to obtain
reasonable agreement with the experimental data.

\begin{figure}[ht]
\begin{center}
\vspace{1cm}
\includegraphics[width=8cm]{./Ic_L_new.eps}\end{center}
\caption{A close-up of the strength data for $D=0.6$ and various $a_0$ and $L$.
The inset compares fracture strength in unnotched samples with that 
predicted by Eq.~(\ref{baz}) for $a_0=0$.}
\label{fig3}
\end{figure}

\begin{figure}[ht]
\begin{center}
\includegraphics[width=8cm]{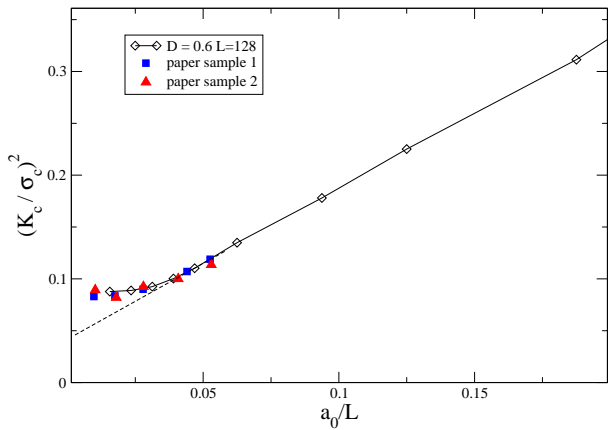}\end{center}
\caption{Comparison of numerical results with the experimental data
for two kinds of paper. Figure presents size effect for small notch sizes using
a RFM simulation of a system of size $L=128$ and a disorder of $D=0.6$.}
\label{fig4}
\end{figure}

The observations from Figs.~\ref{fig2} and ~\ref{fig3} can be summarized into a single scaling
theory by noticing that there must be a scale $a_c$ above which the LEFM
holds and $\sigma_c$ follows Eq.~(\ref{baz}). For $a_0\ll a_c$ the
strength scaling deviates significantly from Eq.~(\ref{baz}) and saturates to a value 
that depends only on disorder and the sample size, $\sigma(L,D)$. This is the 
strength of the unnotched system of size $L$ and disorder $D$. The cross-over from 
LEFM scaling (Eq.~(\ref{baz})) occurs at a notch size $a_c$ 
which can be obtained from Eq.~(\ref{baz}) as
$1/\sigma(L,D)^2 \simeq (a_c+\xi)/K_c^2$.
It is possible to describe this crossover by the following
scaling form, valid for all $a_0$ such that
\begin{equation}
\frac{K_c^2}{\sigma_c^2} = \xi+a_0 f(a_c/a_0)
\label{fundamental}
\end{equation}
where the scaling function $f(y)$ fulfills the limits
\begin{eqnarray}
        f(y)\simeq
        \left\{ \begin{array}{ll}
        1  & \mbox{if \quad $y \ll 1$}  \\
        y & \mbox{if \quad $y \gg 1$}
\label{scafunc}
                \end{array}
\right.
\label{scafu}
\end{eqnarray}
The length-scale $a_c \simeq (K_c(D) /\sigma(L,D))^2-\xi(D)$
corresponds to a cross-over scale below which material strength is
governed by disorder strength and system size. When the
notch size  exceeds this cross-over length
scale ($a_0>a_c$), fracture is governed by LEFM. 
At fixed $L$, stronger disorder will increase $a_c$ since the 
strength of unnotched samples decays faster than
the decrease in $K_c$. This is understandable since 
the disorder masks more efficiently the stress concentration 
due to the notch. Likewise, at fixed $D$, $a_c$  
increases with increasing $L$.

\begin{figure}[ht]
\begin{center}
\includegraphics[width=8cm]{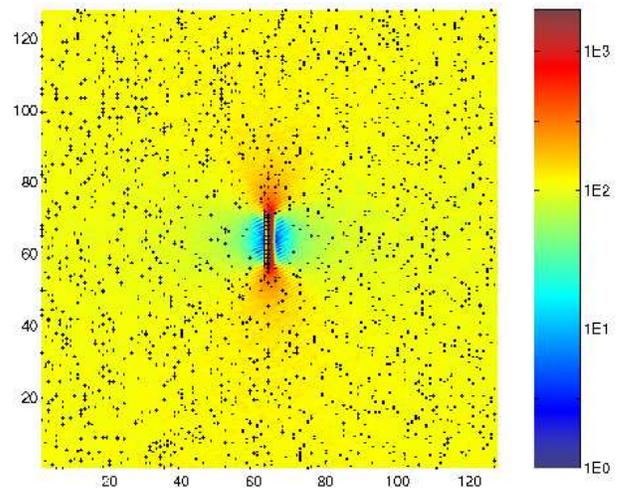}
\end{center}
\caption{The damage (fraction of failed elements or fuses) at
maximum stress $\sigma_c$ for $L=128$ and $D=0.6$. The black markers illustrate the
broken fuses in one single, randomly chosen sample whereas 
the color background represents the damage averaged over 
$N=2000$ number of samples. The colour code indicates the intensity of damage. 
A damage cloud that increases in density is clearly
visible close to the initial notch. The figure demonstrates the
screening of damage due to free crack faces. It is clear that in the 
disordered samples, FPZ is a statistical zone, visible only when damage 
profiles are averaged over many samples.}
\label{fig1}
\end{figure}

\begin{figure}[ht]
\begin{center}
\includegraphics[width=8cm]{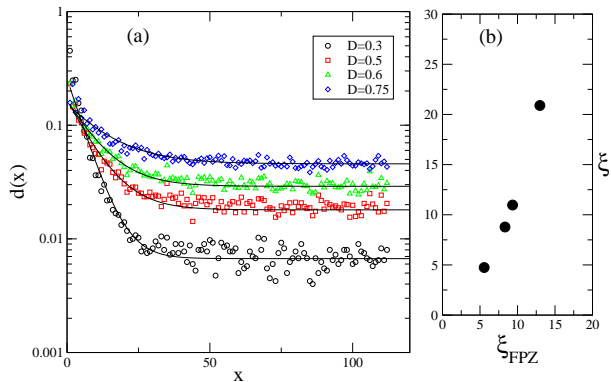}\end{center}
\caption{a): Damage profiles along the crack axis for
various disorders $D$. Damage profiles follow an exponential
decay on a uniform damage background, i.e.,
$d(x)= A+B\exp{(-x/\xi_{FPZ})}$, where $A$ and $B$ are constants and
$x$ is the distance from the crack tip along the crack axis.
b): $\xi$ vs. $\xi_{FPZ}$ for various disorders $D$.}
\label{fig5}
\end{figure}

Simulations of the RFM allow us to access the damage evolution prior
to failure and can thus be used to visualize the development of the
FPZ. As shown in Fig. \ref{fig1}, for a single realization of the
disorder, at {\it maximum stress} we only see diffuse damage,
without apparent localization so that the FPZ can not be observed. 
When we average the damage over different configurations, however,
a clear FPZ emerges in front of the crack tip (see
Fig.~\ref{fig1}). Hence the FPZ should be considered a statistical
concept, visible only when averaging over disorder, while the effect
of the FPZ is nevertheless seen in the size-effect ($\xi$). 
To measure the FPZ size, we consider a projection of the average damage along the crack
direction and obtain a profile that is decaying exponentially
towards a homogeneous background value:
$d(x)=A+B\exp{(-x/\xi_{FPZ})}$ (see Fig.~\ref{fig5}a). We have
analyzed the data for different values of $D$ and checked that the
profiles do not depend on $a_0$ as long as this is not too far from
$a_c$. The LEFM stress intensity factor would indicate a
$1/\sqrt{r}$ -like divergence of the stress at the crack tip. It is
evident that the observed exponential shape of the damage profile $d$ is
in contrast to a $1/\sqrt{r}$ -like decay and should be naturally
interpreted as a screening of the crack tip caused by the disorder.
In Figure \ref{fig5}b, we plot the deduced fracture process size
$\xi_{FPZ}(D)$ against the intrinsic scale $\xi$ that one obtains
from the fits of the strength data to the Eq.~(\ref{baz}) and which
also is an important part of the scaling theory presented in
Eq.~(\ref{fundamental}). It can be seen that these are linearly
proportional indicating that $\xi$ is indeed a direct measure of the
FPZ size. Notice that an exponential damage zone has been indeed measured 
in paper samples and the corresponding lengthscale was compared with the
one obtained from Eq.~(\ref{baz}) \cite{kett}.

In conclusion, we have resorted to simulations of statistical
fracture models to analyze the problem of the size-effect in the
failure of materials. For large notches, the
simulations recover the expected scaling of LEFM
\cite{bazant04b} and allow to relate the effective
FPZ size $\xi$ to the actual average damage profiles. As the notch
size is decreased we observe a crossover at a novel scale $a_c$
to a disorder-dominated
size-dependent regime that is not described by LEFM and is
furthermore seen in experiments. All the regimes are summarized in a
generalized scaling expression (Eq.~(\ref{fundamental})) for the strength 
of disordered media.

{\bf Acknowledgments -} MJA would like to acknowledge the
support of the Center of Excellence -program of the Academy of
Finland, discussions with Dr. R. Wath\'en and the access to the
data of Ref. \cite{wathen03}. MJA and SZ gratefully 
thank the financial support of the European Commissions 
NEST Pathfinder programme TRIGS under contract NEST-2005-PATH-COM-043386.
PKKVN acknowledges support from Mathematical, Information and 
Computational Sciences Division, Office of Advanced Scientific 
Computing Research, U.S. Department of Energy
under contract number DE-AC05-00OR22725 with UT-Battelle, LLC. 
We are grateful M. Gr\"ohn and CSC, Finnish IT Center for 
Science, for assistance.



\begin{thebibliography}{18}
\expandafter\ifx\csname natexlab\endcsname\relax\def\natexlab#1{#1}\fi
\expandafter\ifx\csname bibnamefont\endcsname\relax
  \def\bibnamefont#1{#1}\fi
\expandafter\ifx\csname bibfnamefont\endcsname\relax
  \def\bibfnamefont#1{#1}\fi
\expandafter\ifx\csname citenamefont\endcsname\relax
  \def\citenamefont#1{#1}\fi
\expandafter\ifx\csname url\endcsname\relax
  \def\url#1{\texttt{#1}}\fi
\expandafter\ifx\csname urlprefix\endcsname\relax\def\urlprefix{URL }\fi
\providecommand{\bibinfo}[2]{#2}
\providecommand{\eprint}[2][]{\url{#2}}

\bibitem[{\citenamefont{da~Vinci}(Hoepli Milano 1940)}]{leonardo40}
\bibinfo{author}{\bibfnamefont{L.}~\bibnamefont{da~Vinci}},
  \emph{\bibinfo{title}{I libri di Meccanica}} (\bibinfo{year}{Hoepli Milano
  1940}).

\bibitem[{\citenamefont{Gumbel}(2004)}]{gumbel}
\bibinfo{author}{\bibfnamefont{E.~J.} \bibnamefont{Gumbel}},
  \emph{\bibinfo{title}{Statistics of Extremes}} (\bibinfo{publisher}{Columbia
  University Press, New York}, \bibinfo{year}{2004}).

\bibitem[{\citenamefont{Bazant and Planas}(1997)}]{bazantbook}
\bibinfo{author}{\bibfnamefont{Z.~P.} \bibnamefont{Bazant}} \bibnamefont{and}
  \bibinfo{author}{\bibfnamefont{J.}~\bibnamefont{Planas}},
  \emph{\bibinfo{title}{Fracture and Size Effect in Concrete and Other
  Quasibrittle Materials}} (\bibinfo{publisher}{CRC Press, Boca Raton, USA},
  \bibinfo{year}{1997}).

\bibitem[{\citenamefont{Bazant}(2004)}]{bazant04b}
\bibinfo{author}{\bibfnamefont{Z.~P.} \bibnamefont{Bazant}},
  \bibinfo{journal}{PNAS} \textbf{\bibinfo{volume}{101}},
  \bibinfo{pages}{13400} (\bibinfo{year}{2004}).

\bibitem[{\citenamefont{Hu and Wittmann}(1992)}]{hu92}
\bibinfo{author}{\bibfnamefont{X.}~\bibnamefont{Hu}} \bibnamefont{and}
  \bibinfo{author}{\bibfnamefont{F.}~\bibnamefont{Wittmann}},
  \bibinfo{journal}{Materials Struct.} \textbf{\bibinfo{volume}{25}},
  \bibinfo{pages}{319} (\bibinfo{year}{1992}).

\bibitem[{\citenamefont{Karihaloo}(1999)}]{karihaloo99}
\bibinfo{author}{\bibfnamefont{B.}~\bibnamefont{Karihaloo}},
  \bibinfo{journal}{Int. J. Fracture} \textbf{\bibinfo{volume}{95}},
  \bibinfo{pages}{379} (\bibinfo{year}{1999}).

\bibitem[{\citenamefont{Morel et~al.}(2000)\citenamefont{Morel, Schmittbuhl,
  Bouchaud, and Valentin}}]{morel00}
\bibinfo{author}{\bibfnamefont{S.}~\bibnamefont{Morel}},
  \bibinfo{author}{\bibfnamefont{J.}~\bibnamefont{Schmittbuhl}},
  \bibinfo{author}{\bibfnamefont{E.}~\bibnamefont{Bouchaud}}, \bibnamefont{and}
  \bibinfo{author}{\bibfnamefont{G.}~\bibnamefont{Valentin}},
  \bibinfo{journal}{Phys. Rev. Lett.} \textbf{\bibinfo{volume}{85}},
  \bibinfo{pages}{1678} (\bibinfo{year}{2000}).

\bibitem[{\citenamefont{Morel et~al.}(2002)\citenamefont{Morel, Bouchaud, and
  Valentin}}]{morel02}
\bibinfo{author}{\bibfnamefont{S.}~\bibnamefont{Morel}},
  \bibinfo{author}{\bibfnamefont{E.}~\bibnamefont{Bouchaud}}, \bibnamefont{and}
  \bibinfo{author}{\bibfnamefont{G.}~\bibnamefont{Valentin}},
  \bibinfo{journal}{Phys. Rev. B} \textbf{\bibinfo{volume}{65}},
  \bibinfo{pages}{104101} (\bibinfo{year}{2002}).

\bibitem[{\citenamefont{Carpinteri and Pugno}(2005)}]{carpinteri05}
\bibinfo{author}{\bibfnamefont{A.}~\bibnamefont{Carpinteri}} \bibnamefont{and}
  \bibinfo{author}{\bibfnamefont{N.}~\bibnamefont{Pugno}},
  \bibinfo{journal}{Nature Mat.} \textbf{\bibinfo{volume}{4}},
  \bibinfo{pages}{421} (\bibinfo{year}{2005}).

\bibitem[{\citenamefont{Griffith}(1920)}]{griffith20}
\bibinfo{author}{\bibfnamefont{A.~A.} \bibnamefont{Griffith}},
  \bibinfo{journal}{Trans. Roy. Soc. (london) A}
  \textbf{\bibinfo{volume}{221}}, \bibinfo{pages}{163} (\bibinfo{year}{1920}).

\bibitem[{\citenamefont{de~Arcangelis et~al.}(1985)\citenamefont{de~Arcangelis,
  Redner, and Herrmann}}]{deArcangelis85}
\bibinfo{author}{\bibfnamefont{L.}~\bibnamefont{de~Arcangelis}},
  \bibinfo{author}{\bibfnamefont{S.}~\bibnamefont{Redner}}, \bibnamefont{and}
  \bibinfo{author}{\bibfnamefont{H.~J.} \bibnamefont{Herrmann}},
  \bibinfo{journal}{J. Phys. (Paris) Lett.}
  \textbf{\bibinfo{volume}{46}}, \bibinfo{pages}{585}
  (\bibinfo{year}{1985}).

\bibitem[{\citenamefont{Alava et~al.}(2006)\citenamefont{Alava, Nukala, and
  Zapperi}}]{alava06}
\bibinfo{author}{\bibfnamefont{M.~J.} \bibnamefont{Alava}},
  \bibinfo{author}{\bibfnamefont{P.}~\bibnamefont{Nukala}}, \bibnamefont{and}
  \bibinfo{author}{\bibfnamefont{S.}~\bibnamefont{Zapperi}},
  \bibinfo{journal}{Adv. Phys.} \textbf{\bibinfo{volume}{55}},
  \bibinfo{pages}{349} (\bibinfo{year}{2006}).

\bibitem[{\citenamefont{Nukala et~al.}(2005)\citenamefont{Nukala, Zapperi, and
  Simunovic}}]{nukala05}
\bibinfo{author}{\bibfnamefont{P.~K. V.~V.} \bibnamefont{Nukala}},
  \bibinfo{author}{\bibfnamefont{S.}~\bibnamefont{Zapperi}}, \bibnamefont{and}
  \bibinfo{author}{\bibfnamefont{S.}~\bibnamefont{Simunovic}},
  \bibinfo{journal}{Phys. Rev. E} \textbf{\bibinfo{volume}{71}},
  \bibinfo{pages}{066106} (\bibinfo{year}{2005}).

\bibitem[{\citenamefont{Lilliu and van Mier}(2003)}]{lilliu03}
\bibinfo{author}{\bibfnamefont{G.}~\bibnamefont{Lilliu}} \bibnamefont{and}
  \bibinfo{author}{\bibfnamefont{J.~G.~M.} \bibnamefont{van Mier}},
  \bibinfo{journal}{Eng. Fract. Mech.} \textbf{\bibinfo{volume}{70}},
  \bibinfo{pages}{927} (\bibinfo{year}{2003}).

\bibitem[{\citenamefont{Duxbury et~al.}(1986)\citenamefont{Duxbury, Beale, and
  Leath}}]{duxbury86}
\bibinfo{author}{\bibfnamefont{P.~M.} \bibnamefont{Duxbury}},
  \bibinfo{author}{\bibfnamefont{P.~D.} \bibnamefont{Beale}}, \bibnamefont{and}
  \bibinfo{author}{\bibfnamefont{P.~L.} \bibnamefont{Leath}},
  \bibinfo{journal}{Phys. Rev. Lett.} \textbf{\bibinfo{volume}{57}},
  \bibinfo{pages}{1052} (\bibinfo{year}{1986}).

\bibitem[{\citenamefont{Weibull}(1939)}]{weibull39}
\bibinfo{author}{\bibfnamefont{W.}~\bibnamefont{Weibull}},
  \emph{\bibinfo{title}{A statistical theory of the strength of materials}}
  (\bibinfo{address}{Stockholm}, \bibinfo{year}{1939}).

\bibitem[{\citenamefont{Wath\'en}(2003)}]{wathen03}
\bibinfo{author}{\bibfnamefont{R.}~\bibnamefont{Wath\'en}},
  \emph{\bibinfo{title}{Lic. thesis}} (\bibinfo{address}{HUT, Espoo, Finland},
  \bibinfo{year}{2003}).

\bibitem[{\citenamefont{Kettunen and Niskanen}(2000)}]{kett}
\bibinfo{author}{\bibfnamefont{H.}~\bibnamefont{Kettunen}} \bibnamefont{and}
  \bibinfo{author}{\bibfnamefont{K.}~\bibnamefont{Niskanen}},
  \bibinfo{journal}{J. Pulp Paper Sci.} \textbf{\bibinfo{volume}{28}},
  \bibinfo{pages}{35} (\bibinfo{year}{2000}).

\end{thebibliography}

\end{document}